\documentclass[smallextended]{svjour3}       

\smartqed 
\usepackage{graphicx}
\graphicspath{{./Figs/}}
\DeclareGraphicsExtensions{.pdf,.jpeg,.png,.eps}

\usepackage{booktabs}
\usepackage{url}
\usepackage{multirow}
\usepackage{siunitx} \usepackage{caption}
\usepackage{subcaption}
\usepackage{subcaption,amsfonts,dcolumn}
\usepackage{xspace}
\usepackage[sort,compress]{natbib}
\usepackage{hyperref}
\usepackage{pbox} 

\usepackage{seqsplit}

\usepackage{tabularx}
\usepackage{enumerate}
\usepackage[most]{tcolorbox}
\definecolor{custom-gray}{cmyk}{0, 0, 0, 0.7, 1.00}
\newtcolorbox{Summary}[2][]{
top=0.15in,
fonttitle=\bfseries,
colbacktitle=custom-gray,
colback=gray!5,
colframe=gray!40!black,
enhanced,
attach boxed title to top left={xshift=1.5em,yshift=-\tcboxedtitleheight/2},
boxed title style={size=small,colback=custom-gray},
drop shadow={black!50!white},
title=#2,#1}

\newcommand{\enquote}[1]{``#1''}

\newcommand{\ecofmt}[1]{\texttt{#1}}
\newcommand{\libfmt}[1]{\textsl{#1}}
\newcommand{\verfmt}[1]{\textit{#1}}
\newcommand{\langfmt}[1]{#1}
\newcommand{\BindFind}{\texttt{BindFind}\xspace}

\begin{document}

\title{Bridging the language gap: an empirical study of bindings for open source machine learning libraries across software package ecosystems}
\titlerunning{Bindings for machine learning libraries across package ecosystems}

\author{
    Hao Li         \and
    Cor-Paul Bezemer
}

\institute{
    Hao Li and Cor-Paul Bezemer  \at
    Analytics of Software, GAmes And Repository Data (ASGAARD) Lab \\
    University of Alberta, Edmonton, Alberta, Canada \\
    \email{\{li.hao, bezemer\}@ualberta.ca}
}

\date{Received: date / Accepted: date}

\maketitle

\begin{abstract}
Open source machine learning (ML) libraries enable developers to integrate advanced ML functionality into their own applications. However, popular ML libraries, such as TensorFlow, are not available natively in all programming languages and software package ecosystems. Hence, developers who wish to use an ML library which is not available in their programming language or ecosystem of choice, may need to resort to using a so-called binding library (or \textit{binding}). Bindings provide support across programming languages and package ecosystems for reusing a host library. For example, the Keras .NET binding provides support for the Keras library in the NuGet (.NET) ecosystem even though the Keras library was written in Python. In this paper, we collect 2,436 cross-ecosystem bindings for 546 ML libraries across 13 software package ecosystems by using an approach called \BindFind, which can automatically identify bindings and link them to their host libraries. Furthermore, we conduct an in-depth study of 133 cross-ecosystem bindings and their development for 40 popular open source ML libraries. Our findings reveal that the majority of ML library bindings are maintained by the community, with npm being the most popular ecosystem for these bindings. Our study also indicates that most bindings cover only a limited range of the host library's releases, often experience considerable delays in supporting new releases, and have widespread technical lag. Our findings highlight key factors to consider for developers integrating bindings for ML libraries and open avenues for researchers to further investigate bindings in software package ecosystems.
\end{abstract}

\newcommand{\rqone}{What is the prevalence of bindings for ML libraries in software package ecosystems, and how effective is \BindFind in identifying them?}
\newcommand{\rqtwo}{How are ML libraries and their bindings distributed across ecosystems?}
\newcommand{\rqthree}{How are cross-ecosystem bindings for popular ML libraries maintained?}

\newcommand{\motivation}{\textit{Motivation. }}
\newcommand{\approach}{\medskip\noindent\textit{Approach. }}
\newcommand{\findings}{\medskip\noindent\textit{Findings. }}

\section{Introduction}\label{sec:introduction}

Machine learning~(ML) has become extremely popular in the last decade. Nowadays, there exist many ML applications in our daily lives, such as email spam filters, recommendation systems, and voice assistants. To provide ML features in an application, most developers rely on well-developed open source ML libraries, such as \libfmt{TensorFlow}~\citep{abadi_tensorflow_2016} or \libfmt{PyTorch}~\citep{pytorch}. These open source ML libraries provide easy-to-use interfaces for software developers to use ML techniques in their projects. However, these libraries often target only one programming language and publish to one software package ecosystem. For example, \libfmt{scikit-learn}~\citep{scikit-learn}, a popular ML library which provides various ML algorithms, is written in \langfmt{Python} and publishes to \ecofmt{PyPI}. Thus, \langfmt{Python} developers can directly use the published \libfmt{scikit-learn} package through \ecofmt{PyPI} but developers in other programming languages cannot use this library as easily.

Reusing existing libraries is a common practice in open source communities~\citep{zaimi_reuse_2015, heinemann_reuse_2011}, as it can provide economic benefits by improving software quality and development productivity~\citep{mohagheghi2007reuse, barros2018reuse}. However, reusing libraries across programming languages is challenging due to programming language barriers. There exist several workarounds that allow a developer to use a library that was not written in their preferred language. First, they could choose an alternative but similar library that is written in their preferred language. However, such a similar library may not exist, and even if it does, it may only provide a subset of the required functionality. Another workaround is to develop the library from scratch, but this approach is error-prone and requires a large amount of work. Given the complexity of most ML libraries which require extensive expertise for development and maintenance, developing ML libraries from scratch in different programming languages is often impractical. 

Bindings have emerged as a practical solution for reusing libraries across different programming languages. These bindings can act as wrapper libraries that convert interface calls, handle data type translations, and manage resource allocations. They typically rely on foreign function interfaces~(FFIs) or specialized tools to interact with the original library, referred to as the \textit{host library}. For example, the \langfmt{Rust} programming language provides an FFI\footnote{\url{https://doc.rust-lang.org/rust-by-example/std_misc/ffi.html}} to interact with \langfmt{C} libraries, while tools like SWIG (Simplified Wrapper and Interface Generator)\footnote{\url{https://www.swig.org}} support calling \langfmt{C/C++} libraries from various languages such as \langfmt{Python} and \langfmt{JavaScript}. Through bindings, developers can reuse the functionality of the host libraries without having to develop them from scratch.

Given the unique challenges and importance of ML libraries, it is crucial to investigate the prevalence and characteristics of bindings specifically in this context. To enable large-scale analysis of bindings and their host libraries, we introduce an approach called \BindFind for automatically identifying bindings and extracting the names of the host libraries, which we refer to as \textit{host names}. Using \BindFind, we conduct a comprehensive investigation into open source ML libraries and their bindings across the ecosystems of 13 programming languages. Specifically, we seek to answer the following research questions (RQs):

\begin{enumerate}[\bfseries RQ1.]
	\item{
		\textbf{\rqone} 

            Understanding the prevalence of bindings informs developers and researchers about the availability of bindings for ML libraries. Bindings and their host names can be very accurately identified in software package ecosystems using \BindFind. We identified 2,436 bindings for 546 ML libraries across 13 software package ecosystems.

	}
	\item{
		\textbf{\rqtwo}

            Analyzing the distribution of bindings reveals popular ecosystems and combinations, aiding developers with making informed choices about ecosystem support and encouraging cross-ecosystem integration. The most common combination of ecosystems that support an ML library is \ecofmt{npm} with \ecofmt{PyPI}. While \ecofmt{PyPI} is the most popular ecosystem for ML libraries, \ecofmt{npm} is the most popular ecosystem for the bindings of ML libraries.
	}
	\item{
		\textbf{\rqthree}

            Evaluating the maintenance quality of bindings highlights the risks and challenges, guiding better maintenance practices and decisions. Cross-ecosystem bindings often offer low coverage of host library releases, and they suffer from high delays in supporting new releases, and considerable technical lag. The situation is worse for bindings that are not maintained by the official library organizations.
	}
\end{enumerate}

The main contributions of this paper are as follows:

\begin{enumerate}[1.]
    \item The first paper to study bindings for ML libraries within software package ecosystems.
    \item A replication package\footnote{\url{https://doi.org/10.5281/zenodo.12746638}} containing our dataset of 250,668 bindings~(together with their host names) identified by \BindFind. In addition, the replication package includes details on 546 ML libraries and their 2,436 bindings, as well as the results of our analysis in which we matched 3,277 versions of 133 bindings for 40 popular ML libraries~(including 3,785 tags).
    \item A framework for understanding how well the bindings of ML libraries are maintained. Our findings offer a foundation for developers to make informed decisions when selecting bindings.
\end{enumerate}

\paragraph{Paper Organization.} The rest of this paper is organized as follows. Section~\ref{sec:background} gives background information about our study. Section~\ref{sec:relatedwork} discusses related work. Section~\ref{sec:methodology} presents our methodology. Section~\ref{sec:results} presents the findings of our three RQs. Section~\ref{sec:implications} discusses the implications of our findings. Section~\ref{sec:threadstovalidity} outlines threats to the validity of our study. Section~\ref{sec:conclusion} concludes the paper.
 \section{Background}\label{sec:background}

In this section, we give background information about software package ecosystems and cross-ecosystem bindings for ML libraries.

\subsection{Software Package Ecosystems}

Traditionally, developers of open source libraries published their source code in a source code repository like Git~\citep{git}. Developers who wish to use those libraries could then download them directly from the source code repositories. However, developers had to resolve the library's dependencies and build the library manually. To help developers integrate a library more easily, the releases of software libraries can be published to a software package ecosystem. Open source libraries generally select a software package ecosystem to publish their \textit{main package}, making it the official distribution channel for releases. Alternatively, open source libraries might continue to release versions directly through their Git repositories.

Most modern programming languages come with an official package manager and a package registry. This package manager, the package registry and all the packages are the key components of a software package ecosystem. Usually, a package manager helps developers to manage the dependencies of their applications, for example, by downloading a specific version of a dependency when the application is installed. In addition, package managers help developers publish their applications to the software package ecosystem. Most software package ecosystems of programming languages will provide a website for developers to search and browse the information of stored packages. Some examples of software package ecosystems are \ecofmt{Maven} for \langfmt{Java}, \ecofmt{PyPI} for \langfmt{Python}, and \ecofmt{npm} for \langfmt{JavaScript}.

\subsection{Cross-Ecosystem Bindings for ML Libraries}

According to ISO/IEC TR 10182~\citep{ISOIEC_10182_2016}, a language binding is defined as a specification that maps the functional interface of a system facility~(e.g., a software library) to a programming language. In practice, bindings serve as bridges between different programming languages, enabling software written in one language to use libraries developed in another. For example, \libfmt{Tkinter}\footnote{\url{https://docs.python.org/3/library/tkinter}} provides a \langfmt{Python} interface to the well-known \libfmt{Tk} GUI toolkit, which is implemented in \langfmt{C}. In the context of ML, libraries are often tailored to specific languages and ecosystems~\citep{ben_braiek_open-closed_2018}, such as \libfmt{TensorFlow} and \libfmt{PyTorch} mainly target \langfmt{Python} and are distributed via \ecofmt{PyPI}. Through bindings, developers can use these state-of-the-art ML libraries across languages and ecosystems. This can be achieved through various mechanisms, including wrapper libraries that act as intermediaries, software development kits~(SDKs) that bundle libraries with tools for development, or application programming interfaces~(APIs) that facilitate interaction between different software components.

In our study, we use the term \textit{cross-ecosystem bindings} to refer to packages that allow the host library to be utilized in other software package ecosystems. For example, \libfmt{tensorflow} in \ecofmt{PyPI} and \libfmt{tfjs} in \ecofmt{npm}~\citep{tfjswebsite} are cross-ecosystem bindings for the same ML library~(\libfmt{TensorFlow}), even though they have different names and reside in different software package ecosystems. Since \libfmt{tensorflow} in \ecofmt{PyPI} and \libfmt{tfjs} in \ecofmt{npm} are both maintained by the official organization of the \libfmt{TensorFlow} host library, we consider these two bindings as \textit{officially-maintained bindings}. Officially-maintained bindings are those managed by the same organization that develops the host library, based on the management and ownership of the binding's source code repository rather than the funding source. In contrast, \textit{community-maintained bindings} are managed by individuals or organizations that are not directly associated with the host library's official organization. For example, \libfmt{TensorFlow.NET} in \ecofmt{NuGet} is a community-maintained binding since its owner is different from that of \libfmt{TensorFlow}.

Host libraries and their cross-ecosystem bindings do not necessarily follow the same release schedules and/or strategies. For instance, a cross-ecosystem binding may choose to only support a portion of the releases of its host library. Also, after the host library publishes a release, there may be a delay before a cross-ecosystem binding supports that release~(if at all). In addition, cross-ecosystem bindings might lag in version compared to the latest release of the host library.

 \section{Related work}\label{sec:relatedwork}

In this section, we discuss prior empirical studies of ML libraries, and related work on software ecosystems and foreign function interfaces.

\subsection{Empirical Studies of ML Libraries}

The increasing popularity of ML has led to significant research interest in understanding the development, usage patterns, and challenges of ML libraries. However, there is a lack of research on the usage of ML bindings across different programming languages. This is the first paper to focus on bindings for ML libraries, which play a crucial role in bridging different programming languages and enabling the wider adoption of ML libraries. 

\citet{dilhara_s2_2021} conducted a large-scale empirical study, revealing a significant rise in ML library adoption and identifying common usage patterns and challenges.  Further analysis by \citet{gao_characterize_2024} focused on the supply chain structure and evolution of the \libfmt{TensorFlow} and \libfmt{PyTorch} packages within \ecofmt{PyPI}, uncovering domain specialization, community clusters, and the reasons for packages leaving the supply chains. 

Several studies focused on the problems that developers could face when using ML libraries. \citet{islam_what_2019} mined Q\&A of ten ML libraries on \ecofmt{StackOverflow}, and reported that three types of problems occurred frequently (i.e., type mismatch, data cleaning, and parameter selection). \citet{wei_api_2022} introduced FIMAX to improve API recommendations for Python-based ML libraries based on extracted questions from Stack Overflow related to six popular ML libraries. \citet{lei_why_2023} identified seven primary reasons why ML projects built on \libfmt{TensorFlow} and \libfmt{PyTorch} often encounter compatibility issues due to library version changes, causing code to fail in some projects even when using the same library API.

Several researchers have conducted comparison studies of multiple ML libraries. \citet{grichi_impact_2020} compared ten multi-language ML frameworks with ten multi-language traditional systems and reported that maintainers of these ML frameworks need more time to accept or reject a pull request than traditional systems. \citet{guo_empirical_2019} compared the development and deployment processes of four ML libraries under the same configuration for training of the same models. They found that using different ML libraries can lead to different levels of accuracy. \citet{han_empirical_2020} collected projects that depend on \libfmt{PyTorch}, \libfmt{TensorFlow}, and \libfmt{Theano} on GitHub, and observed four frequent applications (i.e., image and video, NLP, model theory, and acceleration). In addition, most projects depend on these three libraries directly instead of transitively.

\subsection{Software Ecosystems}

\enquote{Software ecosystems} are studied from several angles and even using different definitions~\citep{Franco_Bedoya2017, Manikas2016, Mens2014}. The \enquote{software package ecosystems} term covers a subset of the software ecosystems term. In this paper, we study ML libraries that can be found across multiple software package ecosystems. These package ecosystems have formal processes for maintenance and official distribution channels for releases, aligning with the focus of our study. Our paper is the first to focus on cross-ecosystem bindings. Prior research has identified cross-ecosystem packages by finding those sharing a common source code repository~\citep{intertwining_kannee_2023, constantinou_breaking_2018}. These studies examined packages that are maintained within a single repository but are distributed across multiple ecosystems. However, they did not directly address the concept of bindings. While these studies might have unintentionally included some officially-maintained bindings within the same repository, they overlooked both community-maintained bindings and officially-maintained bindings hosted in different \seqsplit{repositories}. Our findings in Section~\ref{sec:RQ1} demonstrate that 94\% of the ML libraries bindings are community-maintained, underscoring a key difference in our work.

Many studies have focused on software package ecosystems. In our prior work~\citep{rust_yank_2021}, we studied the release-level deprecation mechanism in \ecofmt{Cargo}~(\langfmt{Rust}) ecosystem and found that the deprecated releases propagate through the dependency network and lead to broken releases. \citet{german_evolution_2013} mined packages in \ecofmt{CRAN} and reported that most dependencies point to a core set of packages in the ecosystem. This phenomenon was also observed in another active software package ecosystem -- \ecofmt{npm}~\citep{wittern_look_2016}. \citet{cogo_empirical_2019} studied dependency downgrades in \ecofmt{npm} and found three reasons behind the downgrades: defects, unexpected changes, and incompatibilities. \citet{retention_rubygems_npm, socio_ruby_github} studied social aspects in ecosystems and found that the developers are more likely to abandon an ecosystem if they do not participate in the community, and the Ruby ecosystem is being abandoned. \citet{visual_ecos} proposed a model for visualizing dependencies in ecosystems, and show that \ecofmt{CRAN} packages tend to use the latest releases, but \ecofmt{Maven} packages stay with the older versions. \citet{decan_how_2020} investigated package releases in three software package ecosystems and observed that most pre-releases do not become $\ge1.0.0$ releases. Moreover, software package ecosystems have different practices, policies, and tools for handling breaking changes~\citep{breaking_changes_3_ecos, breaking_changes_18_ecos}.

In addition, researchers studied other types of software ecosystems. \citet{ecos_identification, ref_coupling_ecos} proposed a reference coupling method to identify software ecosystems in GitHub as well as the dependencies in the ecosystems. \citet{bitcoin_ecos} identified the Bitcoin software ecosystem in GitHub and assessed it as a healthy ecosystem. Furthermore, many researchers studied the health of software ecosystems~\citep{health_eco_archi_pract, Jansen2014}. \citet{life_death_ecos} investigated four abandoned software ecosystems and observed that all these ecosystems had a successor or their components were adopted by other systems. \citet{apache_inter_dep} found that projects in the Apache ecosystem get updates when the dependencies published releases for breaking changes or bug fixes. \citet{social_ecos} studied social aspects in proprietary mobile software ecosystems and observed that most developers chose a specific ecosystem based on others' recommendation. Researchers also studied information security and business factors in mobile software ecosystems~\citep{mobile_ecos_it_secur, mobile_ecos_busi_factor}.

\subsection{Foreign Function Interfaces}

Foreign function interfaces (FFI) bridge the gap between different programming languages and allow developers to reuse libraries written in other languages. To verify the correctness of existing bindings, \citet{furr_checking_2008} presented a static checking system that analyzes both bindings and their host library. In addition, \citet{lee_jinn_2010} built bug detection tools for the FFI in \langfmt{Jave} and \langfmt{Python} by performing dynamic analysis. \citet{nakata_fault_2011} categorized link models and fault models of FFI and proposed a logging framework to track the information flow for each model.

Moreover, wrapping up a function to call a library from another programming language is not always applicable, \citet{chiba_foreign_2019} proposed a framework based on code migration to solve this problem. Besides writing the codes of FFI manually, \citet{finne_hdirect_1998} used an interface definition language to allow \langfmt{Haskell} to communicate with both \langfmt{C} and \langfmt{COM}. In addition, \citet{reppy_application_2006} developed a tool to generate foreign interfaces for high-level languages to use the libraries written in \langfmt{C}.

 \begin{figure}[t]
	\centering
	\includegraphics[width=\columnwidth]{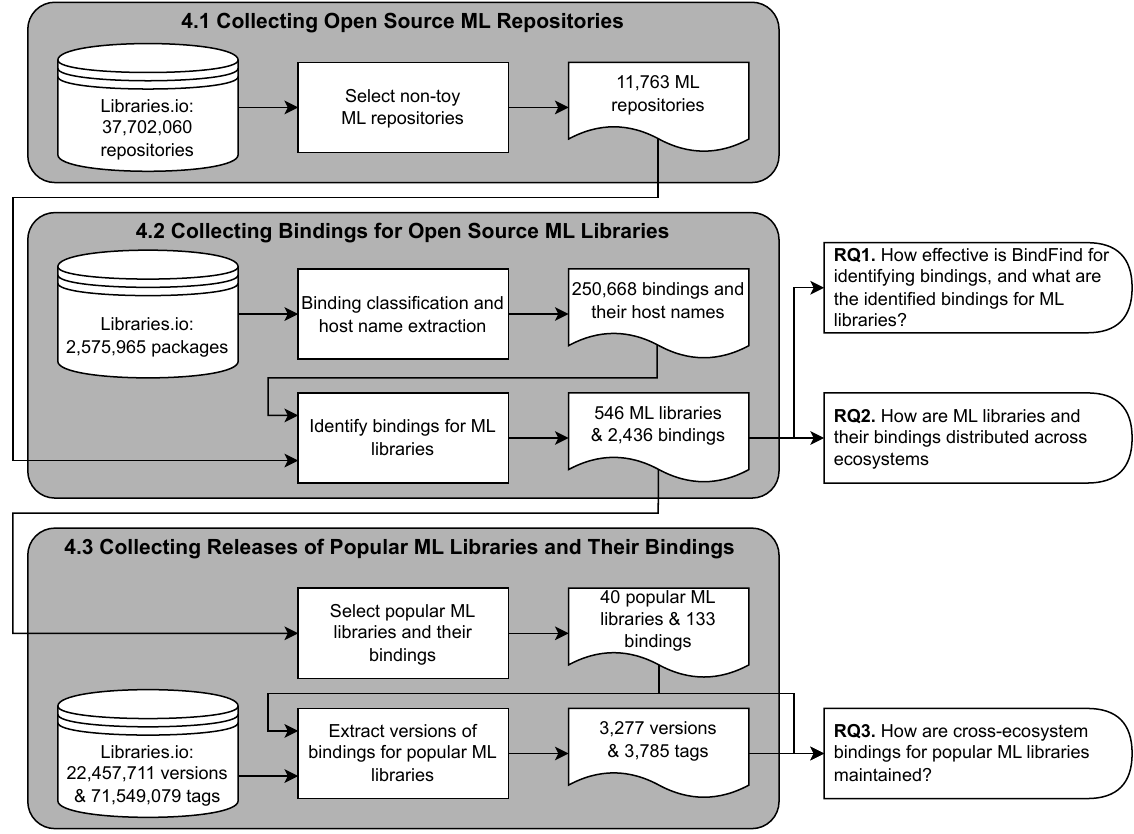}
\caption{Overview of our methodology.}
	\label{fig:methodology}
\end{figure}

\section{Methodology}\label{sec:methodology}

In this section, we introduce the methodology of our study on popular ML libraries and their cross-ecosystem bindings. Figure~\ref{fig:methodology} gives an overview of our methodology.

\begin{table}[t]    
    \caption{Overview of the \ecofmt{Libraries.io} dataset}
    \label{tab:dataset}
    \begin{subtable}{\textwidth}
        \centering
        \caption{Repositories and tags in Git hosts}
        \label{tab:dataset_repos}
        \begin{tabular}{ l r r} 
            \toprule
            \textbf{Platform} & \textbf{\# Repos} & \textbf{\# Tags} \\ 
            \midrule
            GitHub    & 36,567,566 & 58,296,891 \\
            GitLab    & 864,563    & 12,248,518 \\
            Bitbucket & 269,931    & 1,003,670 \\
            \midrule
            \textbf{Total} & 37,702,060 & 71,549,079 \\
        \end{tabular}
    \end{subtable}
    
    \vfill
    
    \begin{subtable}{\textwidth}
        \centering
        \caption{Packages and versions in software package ecosystems}
        \label{tab:dataset_packages}
	\begin{tabular}{ l l r r} 
		\toprule
		\textbf{Ecosystem} & \textbf{Language} & \textbf{\# Packs} & \textbf{\# Versions} \\ 
		\midrule
		\ecofmt{npm}        & \langfmt{JavaScript} & 1,277,221       & 11,400,714      \\
		\ecofmt{Packagist}  & \langfmt{PHP} &  313,575        &  1,766,576      \\
		\ecofmt{PyPI}       & \langfmt{Python} &  232,050        &  1,752,770      \\
		\ecofmt{NuGet}      & \langfmt{C\#} &  199,671        &  2,445,003      \\
		\ecofmt{Maven}      & \langfmt{Java} &  184,890        &  2,799,513      \\
		\ecofmt{RubyGems}   & \langfmt{Ruby} &  161,650        &  1,055,874      \\
		\ecofmt{CocoaPods}  & \langfmt{Objective-C} &   68,085        &    365,782      \\
\ecofmt{CPAN}       & \langfmt{Perl} &   37,496        &    290,847      \\
		\ecofmt{Cargo}      & \langfmt{Rust} &   35,695        &    195,011      \\
		\ecofmt{Clojars}    & \langfmt{Clojure} &   24,295        &    116,945      \\
		\ecofmt{CRAN}       & \langfmt{R} &   16,710        &     94,716      \\
		\ecofmt{Hackage}    & \langfmt{Haskell} &   14,484        &     98,572      \\
\ecofmt{Pub}        & \langfmt{Dart} &   10,143        &     75,388      \\
\midrule
		\textbf{Total} & & 2,575,965 & 22,457,711
	\end{tabular}
    \end{subtable}
\end{table}

\subsection{Collecting Open Source ML Repositories}\label{sec:methodology_1}

We used the \ecofmt{Libraries.io} dataset~\citep{jeremy_katz_2020_3626071} which was updated on January 12, 2020 as our primary data source. This dataset contains information~(e.g., tags, owners, keywords) on 37,702,060 repositories from three prominent Git hosting services. These services host the actual code of open source ML libraries and facilitate version control, developer collaboration, and other functions. Since these hosts are widely used, they may also contain personal projects, documentation, and experimental code. Table~\ref{tab:dataset} outlines the distribution of repositories and tags among these Git hosts.

Following the keyword-matching approach proposed by \citet{ben_braiek_open-closed_2018} to extract ML projects from GitHub, we employed a similar approach to identify ML repositories. This approach involved scanning the \enquote{Description} and \enquote{Keywords} fields of repositories in the dataset for relevant keywords. The keywords included:

\begin{quote}
\enquote{machine learning}, \enquote{deep learning}, \enquote{statistical learning}, \enquote{neural network}, \enquote{supervised learning}, \enquote{unsupervised learning}, \enquote{reinforcement learning}, and \enquote{artificial intelligence.} 
\end{quote}

We crafted regular expressions to accommodate variations in keyword formatting, including the presence of underscores, hyphens, and commas~(e.g., \enquote{machine\_learning}). To ensure we selected representative repositories, we filtered out those with fewer than 5 stars, marked as inactive in the \enquote{Status} field, or indicated as forks. We chose the same cut-off of 5 stars as previous studies~\citep{kochanthara_stars_2022, obrien_stars_2022, bernardo_stars_2024, song_stars_2024} which have shown this threshold to be effective. Ultimately, we extracted 11,763 ML repositories from the dataset.

\subsection{Collecting Bindings for Open Source ML Libraries}\label{sec:methodology_2}

The \ecofmt{Libraries.io} dataset includes information~(e.g., released versions, creation dates, dependencies) on 4,612,919 packages from 38 software package ecosystems. As described in Section~\ref{sec:background}, ML libraries typically have their source code managed on Git hosts and publish packages to these ecosystems. Our analysis focused on 13 selected ecosystems, excluding ecosystems that: (1) focus on a specific domain, such as \ecofmt{Sublime} and \ecofmt{WordPress}, (2) those with a very small number of packages, such as \ecofmt{Shards}~(33 packages) and \ecofmt{PureScript}~(384 packages), and (3) those that do not store information about releases, such as \ecofmt{Go}. Also, we excluded ecosystems that contain duplicated packages of other ecosystems, for example, most packages in \ecofmt{Bower} can be found in \ecofmt{npm}. Table~\ref{tab:dataset} shows the supported programming language, the number of packages, and the number of releases in these 13 ecosystems.

\subsubsection{Binding classification and host name extraction}
\begin{figure}[t]
	\centering
	\includegraphics[width=\columnwidth]{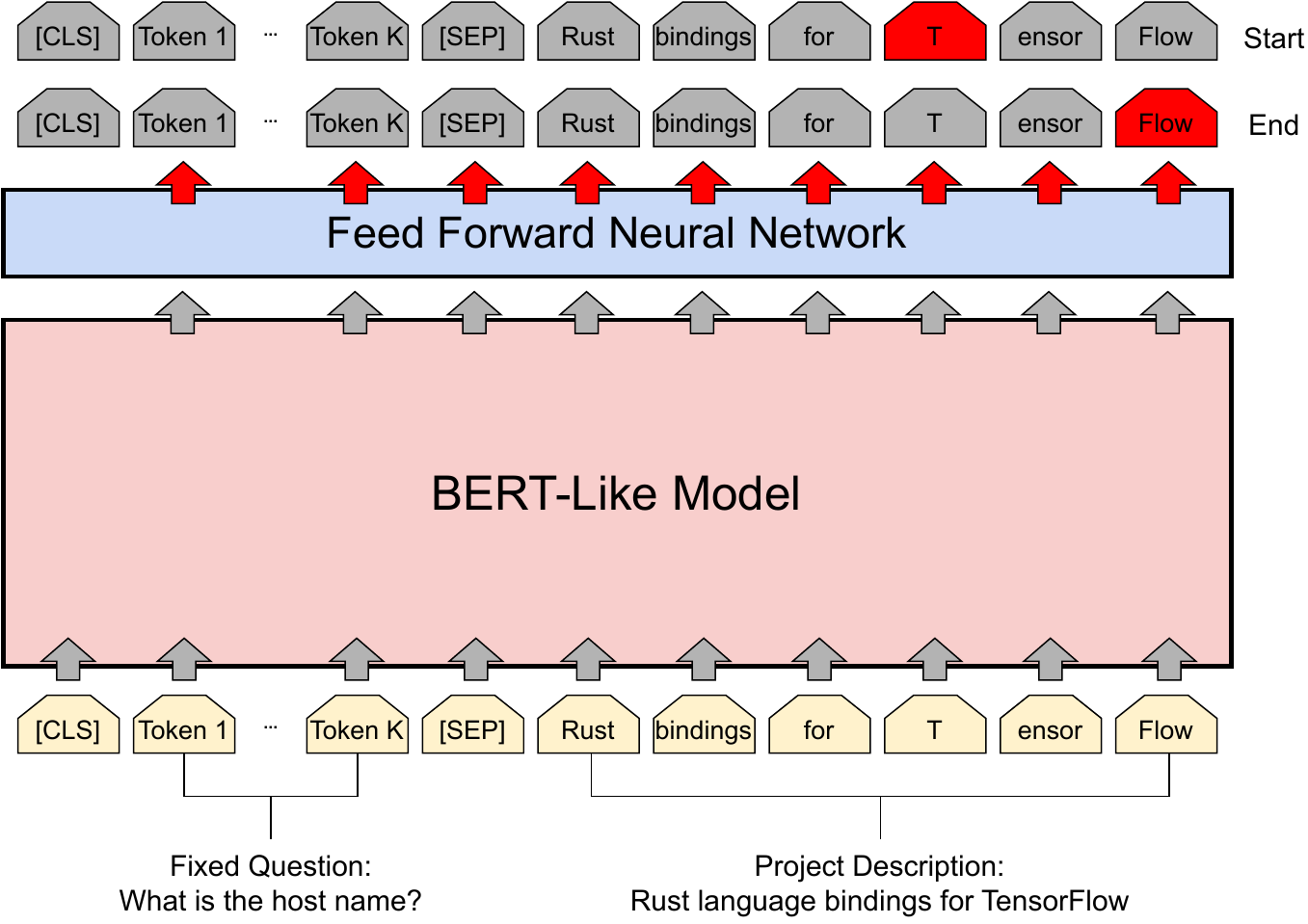}
	\caption{The model structure of \BindFind for binding classification and host name extraction, illustrated using an example.}
	\label{fig:methodology_model}
\end{figure}

To automatically identify bindings in package ecosystems and extract their corresponding host names, we propose \BindFind. \BindFind employs natural language processing~(NLP) techniques, specifically leveraging BERT~(Bidirectional Encoder Representations from Transformers) models~\citep{devlin_bert_2019}. As shown in Figure~\ref{fig:methodology_model}, we conceptualize the problem as an extractive question-answering~(QA) task, akin to the methodology used in the Stanford Question Answering Dataset (SQuAD) v2.0~\citep{rajpurkar_squad2_2018}. In this framework, the description of a package serves as the context for querying the model about the package's host name. This description specifically refers to the textual summary or overview provided by the package ecosystem in the Libraries.io dataset~\citep{jeremy_katz_2020_3626071}. For instance, it can be the \enquote{description} fields in \ecofmt{Maven} and \ecofmt{RubyGems}.\footnote{Examples: \url{https://search.maven.org/artifact/de.vorb/jtesseract/0.0.4/jar} and \url{https://rubygems.org/gems/ruby-opencv/versions/0.0.18}} In other ecosystems, this information might be found in the \enquote{readme} section of \ecofmt{NuGet}\footnote{\url{https://www.nuget.org/packages/Keras.NET}} or the short description of \ecofmt{PyPI}.\footnote{\url{https://pypi.org/project/opencv-python}}

The BERT-like model determines the precise locations~(i.e., the start and end positions) of the host name within the given context. Notably, all the inputs are tokenized before being fed into the model, and the start and end positions refer to the tokens instead of the original input. If the start and end positions point to the \langfmt{[CLS]} token, or if the positions are invalid~(e.g., the start position is after the end position), we conclude that the model did not identify the repository as a binding. This approach adeptly handles both scenarios where questions are answerable and unanswerable, reflecting real-world scenarios where some packages might not be bindings and thus not have a host name to extract.

\paragraph{Studied models.} 
We selected several variations of BERT models, including the original BERT~\citep{devlin_bert_2019}, DistilBERT~\citep{sanh_distilbert_2019}, ALBERT~\citep{lan_albert_2019}, and RoBERTa~\citep{liu_roberta_2019}. These models were chosen for their proven efficacy in QA benchmarks~\citep{rajpurkar_squad1_2016, rajpurkar_squad2_2018}.

\paragraph{Data preparation.} 
We manually labeled 2,546 packages to determine whether they are bindings and, if so, to identify their host names. The dataset contained 2,054 non-bindings~(80.7\%) and 492 bindings~(19.3\%). To ensure a fair evaluation, we balanced the validation and test sets by randomly selecting an equal number of bindings and non-bindings~(50\% each). This approach ensures that performance metrics are not skewed by the imbalanced nature of the dataset. Both the validation and test sets were balanced, each containing 100 samples with an equal distribution of 50 bindings and 50 non-bindings. The remaining 2,346 samples~(392 bindings and 1,954 non-bindings) formed the training set. During the training process, we trained the studied models on the training set and used the validation set for hyperparameter tuning and model selection. The test set was reserved for the final evaluation of model performance.

\paragraph{Evaluation metrics.} 
To assess the effectiveness of \BindFind, we used the \textbf{F1 score}, \textbf{precision}, and \textbf{recall}~\citep{godbole_classification_2004} to evaluate the performance of classifying whether a package is a binding. For the extraction of host names within identified bindings, we applied \textbf{exact match}~(EM) accuracy and (macro-average)~\textbf{F1 score} specific to QA tasks~\citep{rajpurkar_squad2_2018}. EM is a strict metric where any deviation from the exact answer results in a score of 0 for that sample. The F1 score evaluates performance by considering predictions and ground truths as collections of tokens. The reported results are averaged over all of the samples.

\begin{equation}
EM = \frac{{\text{{Number of exact matches}}}}{{\text{{Number of examples that are bindings}}}}
\end{equation}

\begin{equation}
F1 = \frac{2 \times Precision \times Recall}{Precision + Recall}
\end{equation}

where

\begin{equation}
Precision = \frac{\text{Number of accurately identified tokens}}{\text{Number of tokens in the prediction}}
\end{equation}
\begin{equation}
Recall = \frac{\text{Number of accurately identified tokens}}{\text{Number of tokens in the ground truth}}
\end{equation}

\subsubsection{Identify bindings for ML libraries} 
We processed 2,575,965 packages from the 13 studied software package ecosystems and identified a total of 250,668 bindings with their corresponding host names. To identify bindings for the 11,763 ML repositories, we deployed a string-matching algorithm that compared the names of ML repositories with the extracted host names, resulting in 3,360 matches between bindings and 983 ML repositories. We manually reviewed these matches, removing repositories containing only tutorials or experimental code, and filtering out duplicate repositories of the same ML library. After refinement, we identified a total of 2,436 bindings for 546 ML libraries.

It is important to note that some of the 546 ML libraries might publish official packages into ecosystems. These published packages can be either normal packages or bindings~(already identified by \BindFind). For example, \libfmt{PyTorch} publishes its main package \libfmt{torch} in \ecofmt{PyPI},\footnote{\url{https://pypi.org/project/torch}} which is not considered a binding as stated in their documentation~\citep{pytorchdesign}. To study the distribution of ML libraries and their bindings across ecosystems in RQ2, we included officially published packages alongside identified bindings. We used two criteria: (1)~they specified the ML library's Git repository as their source code repository, and (2)~they shared the same homepage URL as the ML library. Using these criteria, we identified 775 packages officially published packages by 202 out of the 546 ML libraries.

\subsection{Collecting Releases of Popular ML Libraries and Their Bindings}\label{sec:methodology_3}

To focus our analysis in RQ3 on widely adopted libraries, we filtered the 546 ML libraries with 2,436 bindings based on the number of stars and selected 127 ML libraries with more than 1,000 stars. Though we acknowledge that stars do not provide a complete picture of real-world usage, they are commonly seen as a proxy for the popularity of a project within the software engineering domain~\citep{stars_borges_2016, stars_fang_2022, stars_han_2019, stars_wolter_2023, stars_xia_2023}. For instance, \libfmt{TensorFlow}'s binding \libfmt{tfjs} has gained over 17,000 stars on GitHub,\footnote{\url{https://github.com/tensorflow/tfjs}} suggesting significant attention from developers. For further analysis, we manually reviewed these libraries and their bindings to perform several refinements. We excluded supporting packages for the actual bindings within the same ecosystem, along with bindings that are either work-in-progress or have only placeholder/invalid releases, e.g., \libfmt{OpenCV}'s binding in \ecofmt{Pub}.\footnote{\url{https://pub.dev/packages/flutter_opencv_plugin}} Following these steps, we obtained a final set of 40 popular ML libraries~(as shown in Table~\ref{tab:bindings_overview}) with 133 bindings.

\begin{table*}[t]
    \centering
    \small
    \caption{Basic information about the popular ML libraries that have cross-ecosystem bindings.}
    \label{tab:bindings_overview}
    \resizebox{\columnwidth}{!}{
    \begin{tabular}{lrrp{0.55\columnwidth}}
        \toprule
        \textbf{ML library} & \textbf{\# Eco} & \textbf{\# Stars} & \textbf{Description} \\
        \midrule
        \libfmt{Alluxio} & 2 & 4,449 & Data orchestration for ML in the cloud \\
        \libfmt{BerryNet} & 1 & 1,150 & Deep learning gateway on Raspberry Pi \\
        \libfmt{bert-as-service} & 2 & 6,379 & Sentence vector mapping with BERT \\
        \libfmt{BigDL} & 2 & 3,177 & Distributed deep learning library for Apache Spark \\
        \libfmt{Bullet} & 2 & 5,798 & Physics simulation for RL \\
        \libfmt{Caffe} & 3 & 29,655 & Deep learning library \\
        \libfmt{CatBoost} & 2 & 3,760 & Gradient Boosting on Decision Trees \\
        \libfmt{Deeplearning4j} & 2 & 11,328 & Deep learning library for Java \\
        \libfmt{DeepSpeech} & 4 & 12,710 & A speech-to-text engine based on TensorFlow \\
        \libfmt{dlib} & 5 & 8,412 & A toolkit for real-world ML applications \\
        \libfmt{DyNet} & 2 & 2,867 & Dynamic Neural Network Toolkit \\
        \libfmt{H2O} & 3 & 4,513 & A platform for distributed ML \\
        \libfmt{ImageAI} & 2 & 4,252 & A library for deep learning and computer vision \\
        \libfmt{Keras} & 3 & 45,995 & A framework to provide human-friendly APIs based on TensorFlow \\
        \libfmt{libpostal} & 6 & 2,171 & An NLP library for address parsing and normalizing \\
        \libfmt{LightGBM} & 3 & 10,242 & A gradient boosting framework \\
        \libfmt{MITIE} & 2 & 1,820 & An NLP library for information extraction \\
        \libfmt{MLflow} & 3 & 5,459 & A ML lifecycle platform \\
        \libfmt{mlpack} & 2 & 3,016 & A library to provide ML algorithms \\
        \libfmt{ncnn} & 1 & 7,930 & Neural network inference \\
        \libfmt{NLTK} & 3 & 8,498 & An NLP library \\
        \libfmt{NNPACK} & 1 & 1,085 & Neural network acceleration \\
        \libfmt{NNVM} & 1 & 1,586 & Compiler for neural nets \\
        \libfmt{ONNX Runtime} & 4 & 1,561 & A runtime engine for ONNX models \\
        \libfmt{OpenAI Gym} & 7 & 19,351 & A toolkit for developing and comparing RL algorithms \\
        \libfmt{OpenCV} & 10 & 41,126 & A computer vision library \\
        \libfmt{OpenFace} & 1 & 12,847 & Face recognition with deep learning \\
        \libfmt{OpenPose} & 1 & 15,532 & Multi-person keypoint detection \\
        \libfmt{Porcupine} & 3 & 1,853 & A library for lightweight wake word detection \\
        \libfmt{PredictionIO} & 8 & 12,226 & A ML server for infrastructure management \\
        \libfmt{PyTorch} & 5 & 35,004 & A ML framework \\
        \libfmt{Rasa} & 2 & 7,436 & A ML framework for automating conversations based on text and voice\\
        \libfmt{scikit-learn} & 3 & 38,756 & A framework to provide ML algorithms \\
        \libfmt{Seldon Core} & 2 & 1,296 & An MLOps framework based on Kubernetes \\
        \libfmt{spaCy} & 2 & 15,161 & An NLP library \\
        \libfmt{TensorFlow} & 11 & 139,939 & A ML framework \\
        \libfmt{Tesseract OCR} & 8 & 32,078 & An OCR engine that uses deep learning \\
        \libfmt{Vowpal Wabbit} & 8 & 6,767 & Techniques to solve interactive ML problems \\
        \libfmt{Weld} & 2 & 1,261 & A library for data analy \\
        \libfmt{XGBoost} & 6 & 17,996 & A gradient boosting framework \\

\bottomrule
    \end{tabular}
    }
\end{table*} 
\section{Results}\label{sec:results}

This section presents the results of our three RQs. For each RQ, we present the motivation, approach, and findings.

\subsection{RQ1: \rqone}\label{sec:RQ1}

\motivation 
Researchers have studied different aspects of open source ML libraries~\citep{dilhara_s2_2021, ben_braiek_open-closed_2018}, but there is a gap in understanding the prevalence of bindings for these libraries. Bindings enable ML library reuse across programming languages, facilitating broader ML integration in diverse projects. However, the absence of comprehensive tools for identifying bindings has been a major obstacle in this area of research. \BindFind fills this gap by offering a method for classifying bindings and extracting host names, thereby enabling future research and practical applications. The findings from this RQ will (1)~inform developers about the availability of bindings for ML libraries, (2)~open new research opportunities for researchers to study bindings in software ecosystems, and (3)~provide a tool for repository hosting platforms~(e.g., GitHub) to discover bindings for assisting developers.

\approach 
We conducted a comparative analysis of various BERT models as described in Section~\ref{sec:methodology_2}. Subsequently, the most effective model was selected and utilized by \BindFind to perform inference across all the packages in the 13 studied software package ecosystems. The distinction between officially-maintained and community-maintained bindings was established through an automatic examination of their association with the host ML libraries. Bindings sharing the same source code repository, organization name, or homepage URL as the host ML library were categorized as officially-maintained. Conversely, those lacking such affiliations were categorized as community-maintained.

\begin{table}[t]
\centering
\caption{Performance comparison of \BindFind with different BERT models in bindings classification and host name extraction on the test set. (U: Uncased; C: Cased; Prec: Precision; Rec: Recall)}
\label{tab:bert_qa}
\begin{tabular}{llcccccc}
\toprule
\multirow{2}{*}{\textbf{Family}}     & \multirow{2}{*}{\textbf{Variant}} & \multicolumn{3}{c}{\textbf{Binding Classification}}                  &  & \multicolumn{2}{c}{\textbf{Host Name Extraction}} \\ \cmidrule(lr){3-5} \cmidrule(lr){7-8}
                            &                          & \textbf{F1}          & \textbf{Prec}      & \textbf{Rec}      &  & \textbf{F1}                  & \textbf{Exact Match}                 \\ \midrule
\multirow{2}{*}{DistilBERT} & Base (U)             & 0.903       & 0.977          & 0.840       &  & 0.806               & 0.780              \\
                            & Base (C)               & 0.936       & 1.000          & 0.880       &  & 0.802               & 0.740              \\ \midrule
\multirow{4}{*}{BERT}       & Base (U)             & 0.925       & 1.000          & 0.860       &  & 0.770               & 0.760              \\
                            & Large (U)            & 0.925       & 1.000          & 0.860       &  & 0.799               & 0.760              \\
                            & Base (C)               & 0.947       & 1.000          & 0.900       &  & 0.803               & 0.760              \\
                            & Large (C)              & 0.936       & 1.000          & 0.880       &  & 0.767               & 0.720              \\ \midrule
\multirow{4}{*}{ALBERT}     & Base                     & 0.918       & 0.938          & 0.900       &  & 0.824               & 0.760              \\
                            & Large                    & 0.925       & 1.000          & 0.860       &  & 0.818               & 0.800              \\
                            & XLarge                   & 0.876       & 1.000          & 0.780       &  & 0.696               & 0.640              \\
                            & XXLarge                  & 0.959       & 0.979          & 0.940       &  & 0.887               & 0.840              \\ \midrule
\multirow{2}{*}{\textbf{RoBERTa}}    & \textbf{Base}            & \textbf{0.970}       & \textbf{0.980}          & \textbf{0.960}       &  & \textbf{0.889}               & \textbf{0.860}              \\
                            & Large           & 0.969       & 1.000          & 0.940       &  & 0.869               & 0.840              \\ \bottomrule
\end{tabular}
\end{table}

\findings
\textbf{5\% of the ML repositories have bindings in software package ecosystems and a vast majority~(94\%) of them are community-maintained bindings.} We found that 5\%~(546 out of 11,763) of the ML repositories are ML libraries with bindings in software package ecosystems. Among the 2,436 identified bindings for these ML libraries, a staggering 94\% (2,292 out of 2,436) of the bindings are maintained by the community. Conversely, only 6\%~(144 out of 2,436) of the bindings are maintained by the official organization. Notably, 58\% (84 out of 144) of these officially-maintained bindings share the same source repository as their host ML libraries, while 42\%~(60 out of 144) are hosted under the same organizational umbrella but in separate repositories. For instance, \libfmt{PyTorch} has officially published a binding~\citep{pytorchmobile} in \ecofmt{Cocoapods} for \langfmt{iOS}, and \libfmt{tfjs-node} serves as an official binding for \libfmt{TensorFlow} within the \ecofmt{NPM} ecosystem~\citep{tfjswebsite}, with its source repository is maintained by the \libfmt{tensorflow} organization on GitHub. Additionally, we observed the transition of some community-maintained bindings to official organizations, such as the \langfmt{Python} binding for \libfmt{OpenCV}~\citep{opencvpythonissue}.

\textbf{\BindFind, powered by the RoBERTa base model, demonstrates exceptional performance in the tasks of binding classification and host name extraction, achieving F1 scores of 0.970 and 0.889 respectively.} As illustrated in Table~\ref{tab:bert_qa}, the RoBERTa base model exhibits superior performance in both the binding classification and host name extraction tasks. Notably, all evaluated models achieved an F1 score above 0.870 in binding classification, indicating a generally high level of accuracy across different architectures. Furthermore, the results demonstrate that larger models do not always guarantee better performance. For instance, the BERT base~(cased) model outperforms its larger variant, and the RoBERTa base model surpasses the RoBERTa large model in both tasks.

\smallskip
\begin{Summary}{Takeaway of RQ1}{}
Using \BindFind, we found that 5\% of the ML repositories have bindings in software package ecosystems. \BindFind demonstrated high effectiveness in identifying bindings (with a 0.970 F1 score) and extracting their host names (with a 0.889 F1 score).
\end{Summary}
 \subsection{RQ2: \rqtwo}\label{sec:RQ2}

\motivation
Existing research shows that Python is the most popular language for ML libraries~\citep{ben_braiek_open-closed_2018}, but it remains unclear which other ecosystems provide good support for ML libraries through bindings. Also, we aim to explore whether certain ecosystems or combinations of ecosystems are favoured by these bindings in this RQ. By investigating the distribution of bindings across ecosystems, we can identify patterns of ecosystem support and cross-ecosystem interaction. The findings from this RQ will (1)~provide developers with insights into ecosystems that provide good support for ML other than \langfmt{Python} and (2)~motivate researchers to further explore the adoption of bindings for ML libraries across ecosystems.

\begin{figure}[t]
	\centering
	\includegraphics[width=\columnwidth]{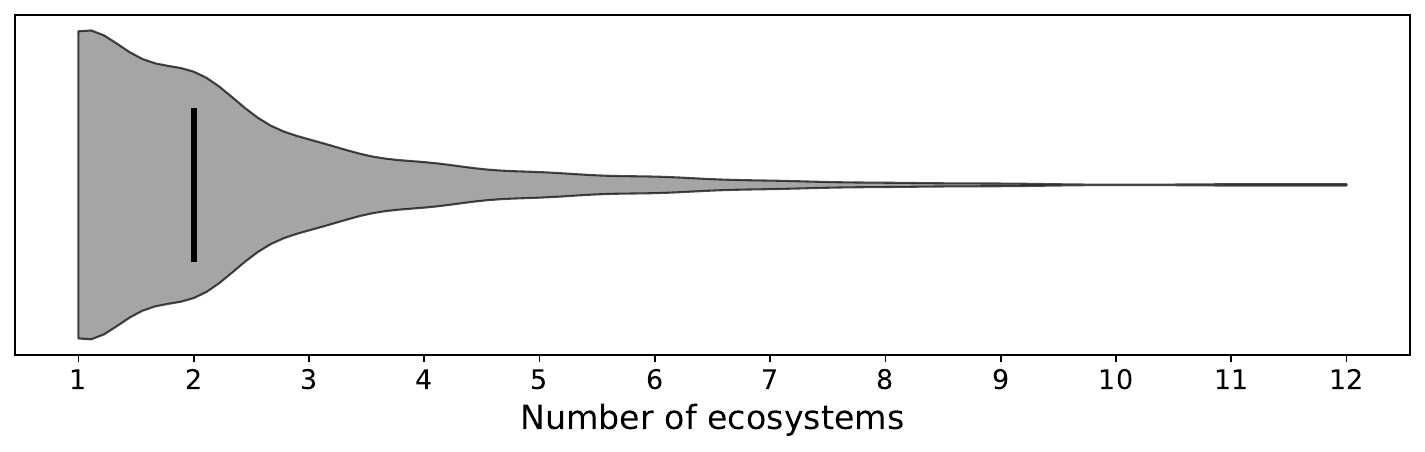}
	\caption{The distribution of the number of software package ecosystems supported by ML libraries with bindings.}
	\label{fig:ml_cross_eco_distribution}
\end{figure}

\approach 
We analyzed 2,436 bindings for 546 ML libraries, with a subset of 202 out of these 546 libraries having officially published packages~(not bindings) across ecosystems~(as detailed in Section~\ref{sec:methodology_2}). Our analysis focused on their distribution across 13 software package ecosystems. To better understand in which combinations of software package ecosystems these libraries reside, we counted the \textit{ecosystem-pairs} for each library. An ecosystem-pair is counted for each pair of ecosystems supporting the same library. For example, if a library is supported in \ecofmt{PyPI}, \ecofmt{npm}, and \ecofmt{NuGet}, we count three ecosystem-pairs: \ecofmt{PyPI}-\ecofmt{npm}, \ecofmt{PyPI}-\ecofmt{NuGet}, and \ecofmt{npm}-\ecofmt{NuGet}. If an ecosystem-pair appears more frequently than others, it implies that those two ecosystems are more likely to be supported together by ML libraries. The frequency of these ecosystem-pairs reveals the prevalence of cross-ecosystem interactions, which is not apparent when considering individual ecosystems alone. Prevalence ecosystem-pairs demonstrate the flexibility developers have when switching between languages to use ML libraries in their projects.

\findings
\textbf{ML libraries with bindings typically span across a median of two software package ecosystems.} Our analysis revealed that ML libraries with bindings are typically supported across a median of two software package ecosystems. As illustrated in Figure~\ref{fig:ml_cross_eco_distribution}, 55\%~(302 out of 546) of the studied ML libraries extend their reach by residing in multiple ecosystems. For example, the NLP library \libfmt{spaCy} has bindings available in both \ecofmt{PyPI} and \ecofmt{npm}. Notably, the library with the broadest ecosystem presence is \libfmt{OpenCV} which is available in 12 different ecosystems, followed by \libfmt{TensorFlow} which is available in 11 ecosystems. Among the ML libraries with bindings that are found in a single ecosystem~(244 out of 546 libraries), 71\%~(174 out of 244) consists of libraries with only community-maintained bindings, lacking official packages or officially-maintained bindings. The remaining cases~(70 out of 244) comprise libraries that have officially-maintained bindings.

\begin{figure}[t]
	\centering
	\includegraphics[width=\columnwidth]{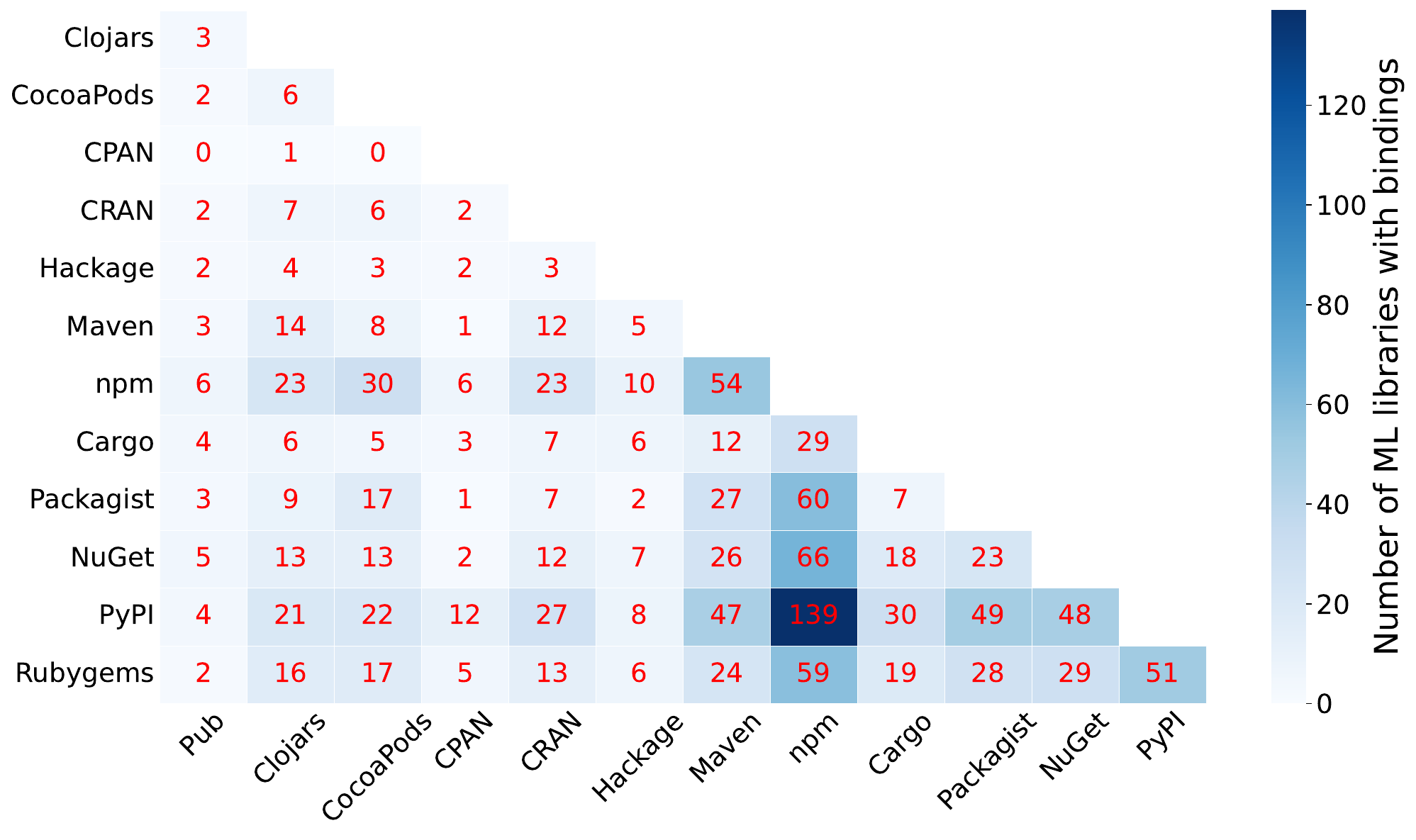}
\caption{Combinations of software package ecosystems in which ML libraries with bindings are available. The elements represent the number of libraries that can be found in both ecosystems~(i.e., ecosystems in the row and column).}
	\label{fig:ml_bindings_pairs}
\end{figure}

\textbf{npm is the leading ecosystem for hosting bindings for ML libraries and the most common combination of bindings is npm with PyPI.} Regarding bindings for ML libraries, we found that 53\%~(292 out of 546) of the libraries have bindings in \ecofmt{npm}. This is closely followed by \ecofmt{PyPI} in which bindings for 41\%~(225 out of 546) of the libraries are hosted. Figure~\ref{fig:ml_bindings_pairs} gives an overview of the ecosystem-pairs of ML libraries with bindings across multiple ecosystems~(302 out of 546). The \ecofmt{PyPI}-\ecofmt{npm} pair is identified as the most prevalent combination, as it is supported by 139 out of 302 ML libraries with bindings. One reason could be that \langfmt{Python} is the most popular language for ML development~\citep{ben_braiek_open-closed_2018} and \langfmt{JavaScript} has been the most commonly used programming language~\citep{stackoverflow_survey_2023}. Hence, there could be a need for ML in \ecofmt{npm}, resulting in more support for such bindings. Other notable ecosystem pairs include \ecofmt{npm}-\ecofmt{Packagist}, \ecofmt{Cargo}-\ecofmt{PyPI}, and \ecofmt{npm}-\ecofmt{Maven}, reflecting a diverse landscape of ML library availability and collaboration.

\textbf{Official organizations behind ML libraries with bindings prefer to focus on a single ecosystem, with PyPI being the most popular choice.} Among ML libraries with bindings, 43\%~(236 out of 546) of them have official packages or officially-maintained bindings. When examining the publication behavior of official organizations behind these 236 ML libraries, we observed that 87\%~(205 out of 236) of the libraries prefer to focus on a single ecosystem, with \ecofmt{PyPI} being the predominant choice~(69\%). This preference aligns with prior research by~\citet{ben_braiek_open-closed_2018}, which reports \langfmt{Python}'s dominance in ML development.

\smallskip
\begin{Summary}{Takeaway of RQ2}{}
55\% of ML libraries with bindings are distributed across at least two software package ecosystems, with \ecofmt{npm} being the most popular choice for publishing these bindings. Moreover, the most popular combination of ecosystems to support is \ecofmt{PyPI} and \ecofmt{npm}.
\end{Summary}

 \subsection{RQ3: \rqthree}\label{sec:RQ3}

\motivation
The popularity of certain ML libraries has led to increased development of bindings across software package ecosystems. However, the development and maintenance practices for these bindings can be different from their host libraries. If a binding only supports a small proportion of the releases of its host library and has a high delay in getting an update, developers relying on this binding may be forced to use outdated versions for extended periods. This situation poses risks such as exposure to bugs and vulnerabilities present in older versions. Without understanding the maintenance quality of these bindings, developers face challenges regarding the integration of bindings. The findings from this RQ will (1)~help developers evaluate the maintenance quality of bindings for ML libraries when selecting bindings, (2)~open new research opportunities to explore factors affecting the maintenance of these bindings, (3)~guide ML package owners in improving support and communication strategies for bindings, and (4)~motive tool builders to create tools to support the maintenance of these bindings.

\begin{figure}[t]
	\centering
	\includegraphics[width=\columnwidth]{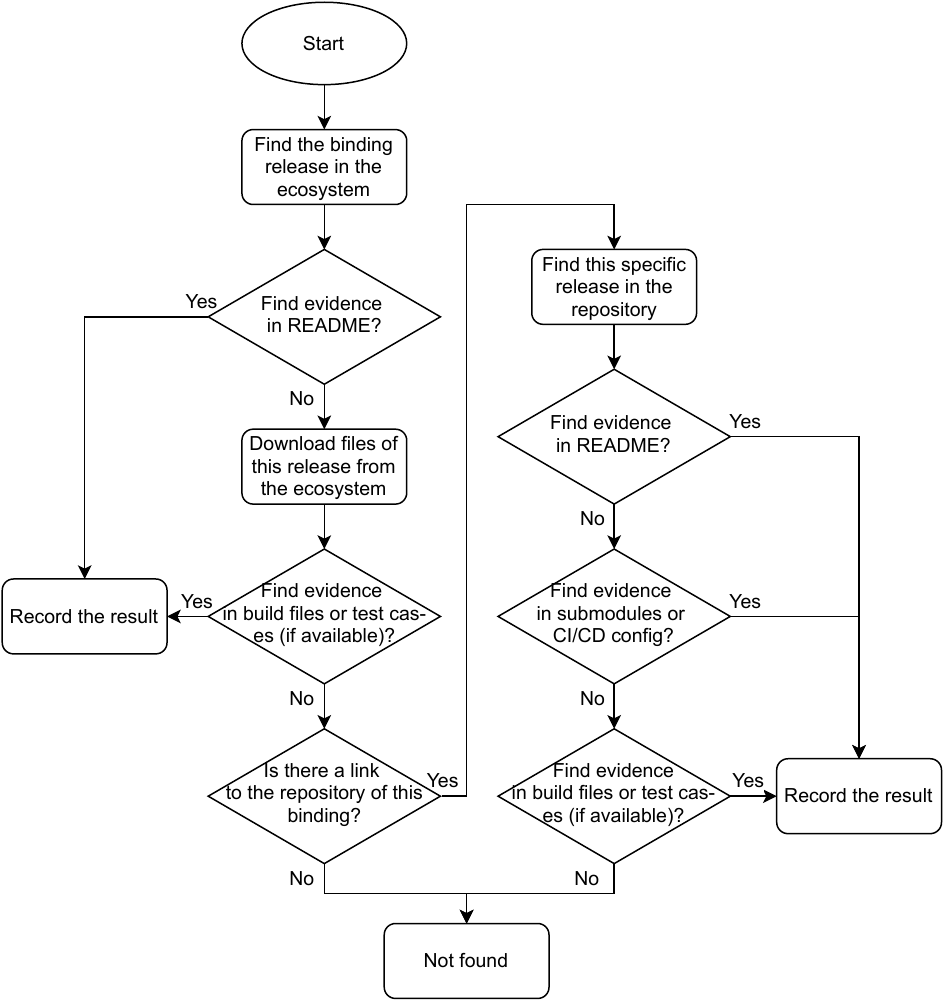}
	\caption{The process of identifying which version of the host library is supported by a specific binding version.}
	\label{fig:rq3_approach}
\end{figure}

\approach 
For libraries that primarily distribute their packages within a single software package ecosystem or designate a specific package as their main release channel, we considered the versions of these packages as the library's releases. For instance, \libfmt{PyTorch} adopts a \enquote{Python First} approach~\citep{pytorchdesign}, positioning its \libfmt{torch} package in \ecofmt{PyPI} not merely as a binding but as the main package. On the other hand, for libraries like \libfmt{OpenPose}, which do not distribute packages through any specific package ecosystem,\footnote{\url{https://github.com/CMU-Perceptual-Computing-Lab/openpose/issues/1250}} we relied on source code repository tags as their releases. This method also applies to bindings that span multiple ecosystems, particularly when no main package is declared. In these cases, the source code repository tags often represent the most reliable source of library releases, such as the tags for \libfmt{XGBoost}.\footnote{\url{https://github.com/dmlc/xgboost}} 

We assume a one-to-one mapping between the binding and the host library versions. Bindings typically follow one of the following approaches: (1)~include the source code of the corresponding host library version, (2)~include the pre-built binary of the corresponding host library version, or (3)~download and compile the corresponding host library version during installation. For instance, the XGBoost binding package in CRAN\footnote{\url{https://cran.r-project.org/web/packages/xgboost/index.html}} follows the first approach by including the source code of a corresponding version of XGBoost. OpenCV's Python binding\footnote{\url{https://github.com/opencv/opencv-python}} follows the second approach by including the built OpenCV binaries for the corresponding version. Finally, TensorFlow's Rust binding\footnote{\url{https://github.com/tensorflow/rust}} follows the third approach by automatically downloading or compiling a corresponding version of TensorFlow. Furthermore, some bindings support developers in installing a specific version of the host library locally and specifying this version when building or installing the binding. Subsequently, we matched all releases of each binding with the corresponding host library releases by searching for evidence of this matching in the following places of the binding releases:

\smallskip\begin{enumerate}[1.]
	\item \textbf{README:} When the supported version of the host library is mentioned explicitly. 
	\item \textbf{Git Submodules~\citep{git}:} When the source code of the supported version of the host library is included as a submodule.
	\item \textbf{Build Files:} When the supported version of the host library that is going to be built for developers is mentioned explicitly, e.g., a binding might indicate the supported version in \textit{CMakeLists.txt} or \textit{Rakefile}.
	\item \textbf{Test Cases:} When the supported version of the host library is verified explicitly by the tests.
	\item \textbf{Configurations of continuous integration or delivery~(CI/CD):} When the supported version of the host library is indicated explicitly in the configuration files, such as \textit{.travis.yml}, to set up the CI/CD environment.
\end{enumerate}

Figure~\ref{fig:rq3_approach} shows an overview of the identification process. For each release of a cross-ecosystem binding, we inspected the README and downloaded the files of this release from the software package ecosystem. If no evidence could be found, we checked out the source code repository of this binding. To locate the corresponding checkpoint~(i.e., a Git commit) of a specific release in the repository, we investigated all tags of the repository and extracted the one that had the same version as the release of the binding. However, some repositories do not have tags for published releases. In this case, we tracked the modification history of the meta-data file~(which stores the version number) to locate the checkpoint. Some examples of meta-data files are \textit{package.json} in \ecofmt{npm}, \textit{setup.py} in \ecofmt{PyPI}, and \textit{pom.xml} in \ecofmt{Maven}. After locating the checkpoint of the specific release, we inspected the files to find out which release of the host library is supported. We performed this process for all releases of cross-ecosystem bindings and matched them with the releases of the host library. The matching results can be found in our replication package~\citep{datarepo}. 

\begin{figure}[t]
	\centering
	\includegraphics[width=\columnwidth]{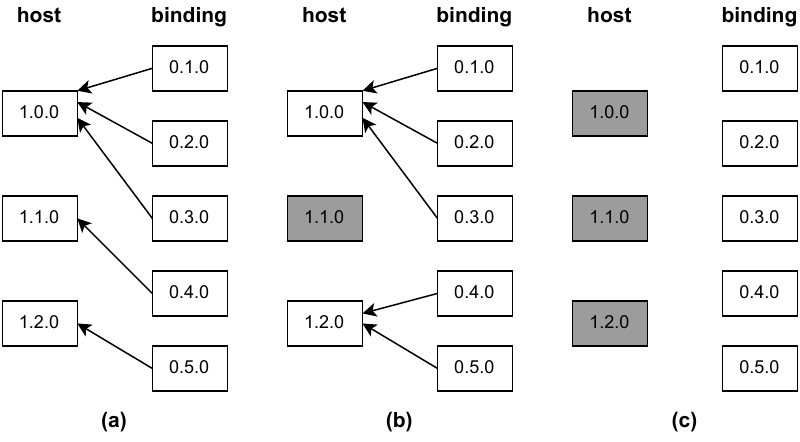}
	\caption{Three examples of matching binding releases and host releases: (a) all host releases are supported by the binding; (b) 2 out of 3 host releases are supported; (c) no host releases are supported.}
	\label{fig:rq3_coverage_example}
\end{figure}

To quantitatively assess the alignment between binding and host library releases, we extracted the \textbf{delay} between the matched releases and the host releases in days and calculated the \textbf{coverage} for each binding $b$ in software package ecosystem $e$ as follows:

\begin{equation}
coverage_{e,b} = \frac{\#matched\_host\_releases_{e,b}}{\#release_{host,b}}
\end{equation}

where the numerator is the number of host releases that are supported by $b$ and the denominator is the number of releases of the host library. We only consider the releases of the host library that were published since the binding started to provide support. The coverage of a binding $b$ will be 100\% if we can find a matched binding release for every host release. If we cannot find any matched release in $b$, the coverage will be 0\%. The coverage metric captures the overall support that a binding offers for an ML library. Figure~\ref{fig:rq3_coverage_example} presents three matching results, the denominators are 3 for all three examples and the numerators are 3, 2, and 0 respectively. Hence, the coverage values for these examples are 100\%, 67\%, and 0\%. In addition, we investigated the \textbf{technical lag}~\citep{technical_lag_2019, technical_lag_2020} of the latest release of each binding. Technical lag occurs when a binding does not support the most recent version of its host library. For example, a major version lag exists when the latest version of a binding supports version \verfmt{1.1.0} of its host library, but the latest available version of the host library is \verfmt{2.1.0}. Similarly, minor and micro version lags exist when there are discrepancies in the minor or micro version numbers respectively.

Next, we compared our findings between officially-maintained bindings and community-maintained bindings by performing the Mann-Whitney U test \citep{Mann1947OnAT} at a significance level of $\alpha=0.05$ to determine whether the differences are significant. Also, we computed Cliff's delta $d$~\citep{Cliff} effect size to quantify the difference. To explain the value of $d$, we use the thresholds which are provided by \citet{Cliff_threshold}:

\begin{equation}
\mathrm{Effect \ size} = 
\left\{
\begin{array}{ll}
	negligible,  & \mathrm{if} \ |d|  \le 0.147 \\
	small,  & \mathrm{if} \ 0.147 < |d|  \le 0.33 \\
	medium,  & \mathrm{if} \ 0.33 < |d|  \le 0.474 \\
	large,  & \mathrm{if} \ 0.474 < |d|  \le 1 \\
\end{array}\right.
\end{equation}

\begin{figure}[t]
	\centering
	\includegraphics[width=\columnwidth]{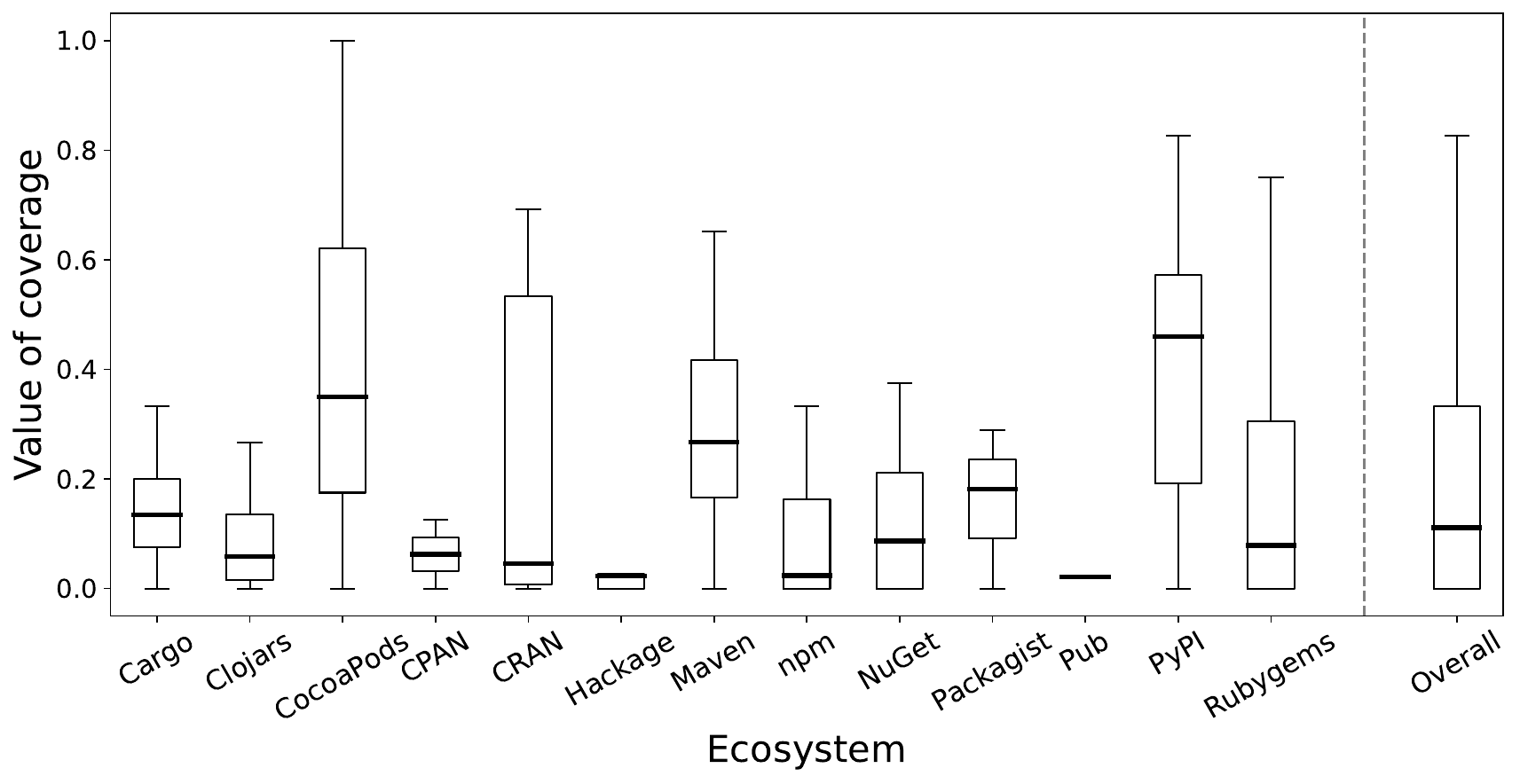}
	\caption{The distributions of the coverage of cross-ecosystem bindings for popular ML libraries across ecosystems.}
	\label{fig:ml_cross_eco_coverage}
\end{figure}

\findings
\textbf{While npm is the most popular ecosystem for cross-ecosystem bindings of popular ML libraries, developers in PyPI are more likely to find a matched release of a cross-ecosystem binding for popular ML host libraries.} We found that 48\% of the popular ML libraries have bindings in \ecofmt{npm} compared to 33\% in \ecofmt{PyPI}. However, Figure~\ref{fig:ml_cross_eco_coverage} shows that \ecofmt{PyPI} has the highest coverage among the 13 studied ecosystems, with a median value of 46\%. In contrast, other ecosystems have relatively small coverage values. We observed that the main reason for the low coverage of these bindings is that they only provide support for a specific subset of the releases from their host libraries. For example, a binding of \libfmt{dlib} in \ecofmt{RubyGems} has 13 releases but only supports 2 out of 50 versions of the host library. Specifically, releases \verfmt{1.0.0} to \verfmt{1.0.3} of this binding support version \verfmt{v18.13} of the host library, then it skipped nine versions~(i.e., \verfmt{v18.14} to \verfmt{v19.3}) of the host library and published 9 releases to support \verfmt{v19.4}. This phenomenon can also be found in other bindings with low coverage. In addition, we noticed that the median coverage values of \ecofmt{CRAN}, \ecofmt{Hackage}, \ecofmt{npm}, and \ecofmt{Pub} are below 5\%. This phenomenon may be attributed to a large portion of community-maintained bindings within these ecosystems~(e.g., 84\% in \ecofmt{npm}), where many bindings stick with a single host library version.

\begin{figure}[t]
	\centering
	\includegraphics[width=\columnwidth]{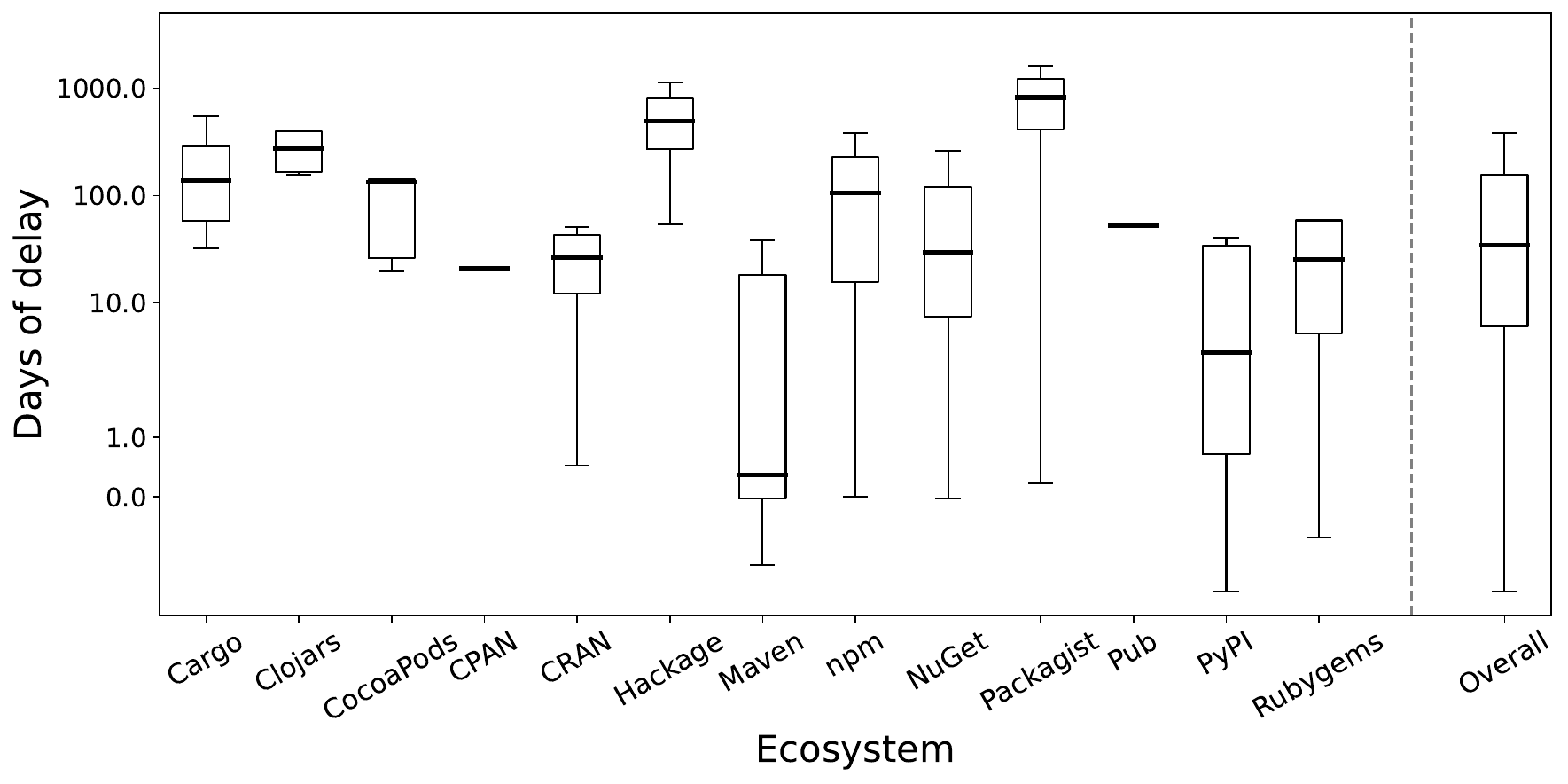}
	\caption{The distributions of the delay between releases of popular ML host libraries and their bindings.}
	\label{fig:ml_cross_eco_delay}
\end{figure}

\textbf{After a release of the ML library was published, their bindings in PyPI and Maven received a corresponding update more quickly than the ones in other software package ecosystems.} Figure~\ref{fig:ml_cross_eco_delay} shows that bindings within the \ecofmt{PyPI} and \ecofmt{Maven} ecosystems tend to be updated to match new releases of ML libraries more swiftly than those in other ecosystems, with median delays of less than 7 days. In contrast, bindings in ecosystems like \ecofmt{Packagist} and \ecofmt{Hackage} may experience median delays exceeding 365 days, indicating a slower pace of alignment with host library updates.

\begin{figure}[t]
	\centering
	\includegraphics[width=\columnwidth]{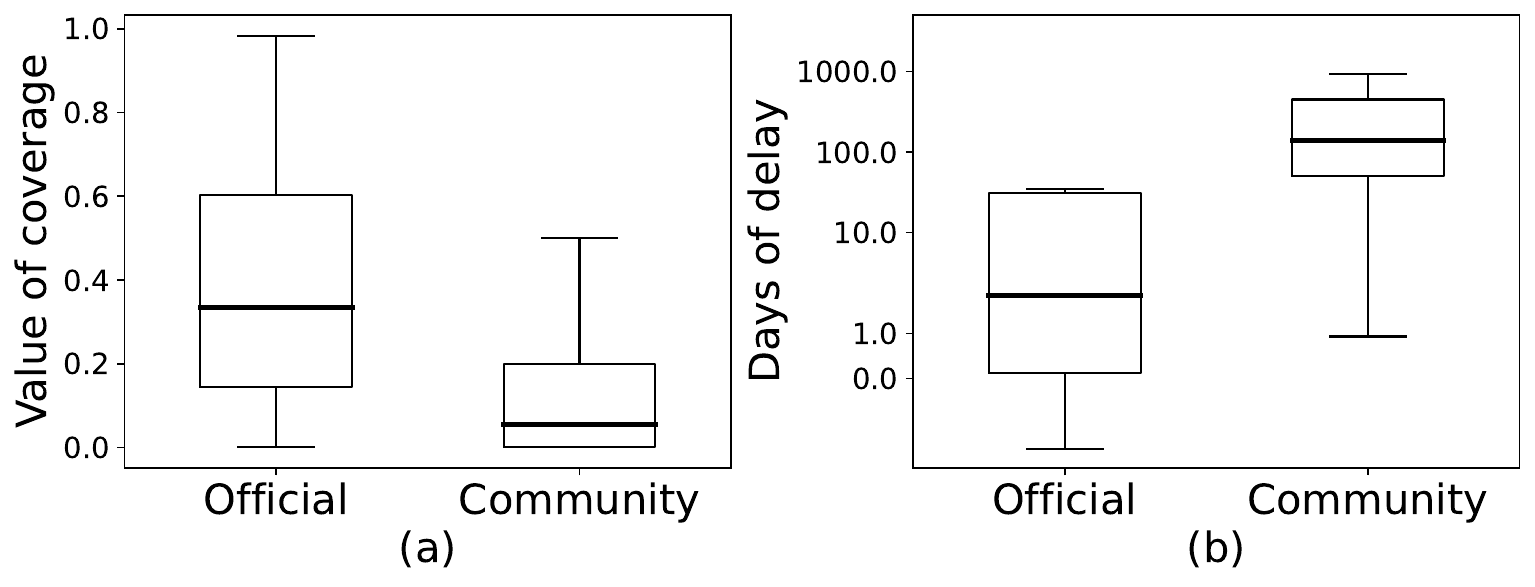}
\caption{Comparisons of the cross-ecosystem bindings for ML libraries which are maintained by the official organization and the community: (a) the distributions of the coverage; (b) the distributions of the delay between a release of the host library and the corresponding release of the binding.}
	\label{fig:ml_cross_eco_official_compare}
\end{figure}

\textbf{Compared to community-maintained bindings for popular ML libraries, the officially-maintained bindings have higher coverage, shorter delays, and smaller technical lags.} Among the 133 bindings for the 40 popular ML libraries, 36 are officially-maintained bindings while 97 are community-maintained bindings~(27\% versus 73\%). Figure~\ref{fig:ml_cross_eco_official_compare} shows that the coverage of community-maintained bindings is mostly below 0.2 and the delays with the host libraries are mostly between 29 and 275 days. In contrast, the officially-maintained bindings have much more coverage (median of 11\%) and less delay (median of 34 days). The Mann-Whitney U test demonstrates significant differences in both coverage and delay distributions between officially-maintained bindings and community-maintained bindings. In addition, the values of Cliff's Delta $d$ are $0.475$ and $-0.819$ respectively, indicating large effect sizes. Furthermore, 45\% of the officially-maintained bindings experience zero technical lag, compared to 24\% of community-maintained bindings. Although officially-maintained bindings show lower proportions of major~(21\% vs. 32\%) and minor~(27\% vs. 40\%) technical lags compared to community-maintained bindings, they have a slightly higher proportion of micro lags~(7\% vs. 4\%).

\textbf{69\% of the cross-ecosystem bindings do not follow the version numbers of their host library and 29\% of cross-ecosystem bindings do not specify the matching information anywhere.} During the process of matching binding releases with their host releases, we noticed that the majority~(92 out of 133) of bindings do not reuse any version number from their host libraries. This misalignment in version numbers could lead to confusion for developers. Furthermore, our manual review process revealed that 29\%~(38 out of 133) of the bindings do not provide explicit matching information. Particularly, the absence of matching information in community-maintained bindings~(89\% of those are without matching information) creates an additional challenge, potentially forcing developers to resort to trial-and-error methods to identify compatible versions.

\smallskip
\begin{Summary}{Takeaway of RQ3}{}
Generally, the release coverage of cross-ecosystem bindings for their host library is low and the delay is large, with PyPI and Maven performing better than other ecosystems. Officially-maintained bindings for popular ML libraries offer significantly better coverage, faster updates, and less technical lag than community-maintained alternatives. However, 69\% of bindings do not follow the version numbers of their host libraries and 29\% of bindings lack explicit version matching information, posing challenges for developers.
\end{Summary}

\section{Implications}\label{sec:implications}

In this section, we discuss the implications of our findings for developers, owners of ML host libraries and their cross-ecosystem bindings, researchers, tool builders, and repository hosting platforms.

\subsection{Implications for Developers}

\textbf{Developers are not always limited to using the same source programming language as a popular ML library when they wish to use this library.} \ecofmt{PyPI} is the most popular choice for ML libraries~\citep{ben_braiek_open-closed_2018} and developers in \ecofmt{PyPI} are more likely to find an ML library for their own projects. However, developers might prefer to start a project in their most familiar ecosystem. Our findings show that 5\% of the ML repositories are libraries that can be found in multiple ecosystems~(Secion~\ref{sec:RQ1}) and their bindings spread across different ecosystems, with \ecofmt{npm} being the most popular choice for bindings~(Secion~\ref{sec:RQ2}). Therefore, developers may find a binding of their desired ML library in the chosen software package ecosystem. We suggest that developers should put the choice of ecosystems into consideration before starting a project.

\medskip\noindent\textbf{Developers should consider the number of supported releases and the delay of getting a corresponding update when choosing the binding for an ML library.} Usually, cross-ecosystem bindings for popular ML libraries do not support all releases of their host library~(Section~\ref{sec:RQ3}). Our findings show that it is not sufficient to look at the number of binding releases only. Bindings with low coverage could publish many releases but only support one or two versions of their host library. If developers are going to adopt such a binding, they should consider that it might not support a needed version in the future. For a binding with high coverage, developers should also consider how long it takes to update for a version of their host library and whether such delays are acceptable to them. By checking the maintenance history of the chosen binding, developers can have an expectation about the binding and consider whether they wish to adopt it or not.

\subsection{Implications for ML Package Owners}

\textbf{Owners of cross-ecosystem bindings for popular ML libraries should explicitly indicate the matching between releases of the binding and releases of the host library.} In Section~\ref{sec:RQ3}, we found that some software package ecosystems have median coverage rates below 5\% for bindings of popular ML host libraries. Such a low coverage indicates that either a small portion of the host releases are supported, or it is not possible to find out which versions are supported~(even after our thorough investigation). In addition, we noticed that only 31\% of cross-ecosystem bindings follow the version numbers of their host library. We recommend that owners of cross-ecosystem bindings use the same version number as their host libraries and indicate that in their README. For example, the binding of \libfmt{mlpack} in \ecofmt{CRAN} mentions that \enquote{the version number of MLPACK is used as the version number of this package}~\citep{RcppMLPACK}. Also, we recommend adding an extra number after the original version number, for example, changing the version number from \enquote{1.2.3} to \enquote{1.2.3.0}. This way, the owner can increase the extra number when fixing bugs in the binding without causing confusion for developers.

\medskip\noindent\textbf{Owners of popular ML host libraries should take notice of the \seqsplit{community-maintained} bindings for their libraries.} 73\% of the cross-ecosystem bindings for popular ML libraries are maintained by the community~(Section~\ref{sec:RQ3}). These community-maintained bindings help developers to use the functionalities from their host library in an ecosystem which the official organization does not support. We recommend that official organizations keep an eye on the community-maintained bindings. For example, the official organization could inform popular community-maintained bindings about important updates, e.g., those that fix security vulnerabilities. We noticed that some libraries list the community-maintained bindings in their README or the official website. Furthermore, we observed that \libfmt{OpenCV} even adopted a popular community-maintained binding~\citep{opencvpythonissue}.

\subsection{Implications for Researchers}

\textbf{Researchers should expand research into bindings across software domains.} The efficacy of \BindFind in identifying and analyzing bindings~(Section~\ref{sec:RQ1}) underscores a rich avenue for further exploration. We suggest researchers apply \BindFind in broader contexts, extending its use to examine areas such as web development frameworks. This extension could provide valuable insights into how bindings enhance software library interoperability across various domains. Researchers can reuse our replication package~\citep{datarepo} which contains a dataset of 250,668 bindings and their host names identified by \BindFind.

\medskip\noindent\textbf{Researchers should further investigate the differences between \seqsplit{officially-maintained} bindings and community-maintained bindings.} Our results show that community-maintained bindings and officially-maintained bindings have different coverages, delays, and technical lags~(Section~\ref{sec:RQ3}). Future studies should investigate what causes these differences. One factor could be simply financial incentives (e.g., because contributors to officially maintained bindings work for the company driving the binding), but there could also be socio-technical factors. For example, developers may be more motivated to contribute to officially-maintained bindings as such contributions are considered more valuable or prestigious. In addition, future studies should investigate how the communication between developers of community-maintained bindings and the host library can be improved.

\medskip\noindent\textbf{Researchers should study automatic matching tools for releases of ecosystem bindings to match with releases of their host library.} It is a complex and tedious task to identify which host library release is supported by an ecosystem binding. Automatic version matching tools for ecosystem bindings can help developers to find a suitable release without going through all the related files of a binding~(like we did in Section~\ref{sec:RQ3}) or trying the releases one by one in their project.

\medskip\noindent\textbf{Researchers should study the impact of using bindings on ML development.} Our replication package~\citep{datarepo} provides information on 546 ML libraries and their bindings. Researchers can utilize this dataset to investigate various aspects of ML development. For instance, prior work~\citep{li_cmpbinding_2024} studied the impact of bindings for \libfmt{TensorFlow} and \libfmt{PyTorch} on software quality in terms of correctness and time cost. Future research can explore how bindings for other ML libraries affect different aspects of software development, such as reliability, maintainability, and efficiency.

\medskip\noindent\textbf{Researchers should explore the adoption of bindings for ML libraries across software ecosystems.} The analysis of the ecosystem distribution of bindings for ML libraries reveals popular ecosystems and ecosystem-pairs~(Section~\ref{sec:RQ2}). These findings suggest a need for further research into the adoption of these bindings across different programming languages and ecosystems. For example, LinkedIn's adoption of TensorFlow’s JavaScript bindings over Python and Java highlights how ecosystem choices can influence the deployment and integration of ML models within existing infrastructure.\footnote{\url{https://blog.tensorflow.org/2022/03/how-linkedin-personalized-performance.html}} By studying such cases, researchers can gain deeper insights into the factors that drive ecosystem preferences and how these decisions affect the success of software projects. Researchers can leverage the dataset provided in our replication package~\citep{datarepo}, which includes detailed information about the distribution of bindings across ecosystems, to conduct their analyses.

\subsection{Implications for Tool Builders}

\medskip\noindent\textbf{Tool builders should create tools to automate and simplify the maintenance of bindings for ML libraries.} Maintaining cross-ecosystem bindings is a challenging task, particularly for community-maintained bindings, indicated by their often lower coverage and higher delays compared to officially-maintained bindings~(Section~\ref{sec:RQ3}). Tool builders should develop tools to monitor new releases of a host library, automatically suggest updates for the bindings to reduce the delay in supporting the latest versions, and check the synchronization of version numbers between bindings and their host libraries. Integrating these tools with CI/CD pipelines can help identify compatibility issues early, improve coverage, and reduce delays for both community and officially maintained bindings.

\subsection{Implications for Repository Hosting Platforms}

\textbf{Repository hosting platforms should integrate \BindFind to discover bindings.} As demonstrated in Section~\ref{sec:RQ1}, \BindFind can accurately identify bindings and extract their host names, making it a valuable tool for repository hosting platforms. For instance, the Tesseract OCR project maintains a manual list of its bindings in the documentation.\footnote{\url{https://tesseract-ocr.github.io/tessdoc/AddOns.html\#tesseract-wrappers}} By integrating \BindFind, repository hosting platforms such as GitHub can automate the discovery of bindings for various host libraries, providing a list of available bindings for developers.

 \section{Threats to Validity}\label{sec:threadstovalidity}

In this section, we discuss the threats to validity of our study about popular ML libraries and their cross-ecosystem bindings.

\subsection{Internal Validity}

We use release coverage as a metric to reason about the maintenance of a binding. A low coverage value does not always mean that a binding has a bad support. For example, a binding can be forward-compatible, thereby making it easier to support newer versions without changing the binding. However, developers have to manually verify whether this binding will work for a specific version of the ML library that they are going to use, as it is not indicated anywhere. A binding can also be backward-compatible, thereby making it support older versions of the host library. However, it is not guaranteed in many cases, as bindings do not typically undergo testing on older versions of the host library, and developers have to verify this themselves.

We assume a one-to-one mapping when matching the release versions in bindings to the ML host libraries. This assumption is based on common practices where bindings either include the exact source code, include pre-built binaries, or download and compile the matching version. However, this assumption does not account for scenarios where a binding supports multiple versions of the host library within a single version.

We classified cross-ecosystem bindings of popular ML libraries into \seqsplit{officially-maintained} bindings and community-maintained bindings. However, some \seqsplit{community-maintained} bindings could be adopted by the official organization and become an officially-maintained binding. In our study, we only consider the latest information and do not take into account the history of ownership. In addition, the owner of a community-maintained binding could be a member of the official organization of the host library. If the binding is not owned by the official organization and there is no evidence in its README, we consider such bindings as community-maintained bindings.

Some software package ecosystems like \ecofmt{PyPI} allow the owner of a package to delete a published release. \ecofmt{Libraries.io} does not record the history of releases of a package, hence, we only consider the information of releases when the dataset was collected. In addition, some ecosystems also support the owner of a package to deprecate a release instead of deleting a release, e.g., \ecofmt{npm} and \ecofmt{Cargo}. In our study, we consider all releases which also include the deprecated ones.

\subsection{External Validity}

In our empirical study, we studied cross-ecosystem bindings for ML libraries. The results of our study might not apply directly to all cross-ecosystem bindings. Still, our methodology can be applied to analyze other cross-ecosystem bindings for libraries in other domains. In addition, we focused on open source ML libraries as many popular ML libraries are open source. Future studies should investigate if our findings hold for proprietary ML libraries.

 \section{Conclusion}\label{sec:conclusion}

In this study, we introduced \BindFind, a novel approach for identifying bindings and extracting host names within software package ecosystems. Applying \BindFind to the \ecofmt{libraries.io} dataset allowed us to gather 2,436 bindings for 546 ML libraries across 13 software package ecosystems. We analyzed the population of bindings for ML libraries and the distribution of these libraries and their bindings across various ecosystems. In addition, we identified 40 popular ML libraries along with their 133 cross-ecosystem bindings, further examining their releases. Based on the data from these popular ML libraries, we conducted an in-depth analysis to investigate the development and maintenance of these cross-ecosystem bindings. We shared the collected data in our replication package~\citep{datarepo}. The most important findings of our study are:

\begin{enumerate}[1.]
    \item \BindFind demonstrates high F1 scores in identifying bindings and extracting host names, indicating its robust capability to assist in the discovery of bindings within software package ecosystems. 
    
    \item \ecofmt{npm} is the most popular software package ecosystem for bindings of ML libraries, with \ecofmt{npm} and \ecofmt{PyPI} emerging as the predominant combination for publishing these bindings.
    
    \item The study highlights significant challenges in the maintenance of cross-ecosystem bindings, including limited release coverage, delays in updating bindings to match new releases, and prevalent technical lag, especially among community-maintained bindings.
 
    \item Most bindings do not follow their host library's version numbers, and many lack clear information on which host library versions they support.
\end{enumerate}

Our findings show that developers who wish to use a popular ML library are not limited to using the programming language the library was written in, as there exist many cross-ecosystem bindings. However, they should carefully check the coverage, delay, and technical lag of these bindings before they commit to using one. In addition, we suggest that maintainers of cross-ecosystem bindings should follow the version number of their host library and add an extra number after it, to account for bug fixes, and make it easier for developers to identify which version of the host library is supported by the binding.

\begin{acknowledgements}
The work described in this paper has been supported by the ECE-Huawei Research Initiative (HERI) at the University of Alberta. \end{acknowledgements}

\section*{Data Availability Statement}
The data generated and analyzed during this study is available in our replication package at \url{https://doi.org/10.5281/zenodo.12746638}.

\bibliographystyle{spbasic}      \bibliography{Bib}   

\end{document}